\documentclass[10pt,a4paper,twocolumn]{article}
\usepackage{geometry}
\usepackage[utf8]{inputenc}
\usepackage{graphicx}
\usepackage[numbers,sort&compress]{natbib}
\usepackage{amsmath}
\usepackage{authblk}
\usepackage[colorlinks=true]{hyperref}
\hypersetup{
    allcolors = {blue},
}
\usepackage{soul}

\usepackage[dvipsnames]{xcolor}

\newcommand{\as}[1]{\textcolor{Orange}{#1}}

\title{\textbf{Rheological softening of metal nanocontacts sheared under oscillatory strains}}

\author[1,2,3]{Ali Khosravi}
\author[1]{Jin Wang}
\author[3,1]{Andrea Silva}
\author[3,1]{Andrea Vanossi}
\author[1,2,3]{Erio Tosatti\thanks{tosatti@sissa.it}}

\affil[1]{International School for Advanced Studies (SISSA), I-34136 Trieste, Italy}
\affil[2]{International Center for Theoretical Physics, I-34151 Trieste, Italy}
\affil[3]{CNR-IOM, Consiglio Nazionale delle Ricerche - Istituto Officina dei Materiali, c/o SISSA,  34136, Trieste, Italy}

\date{\today}

\begin{document}
\maketitle

\begin{abstract}
The way metal interfaces evolve during frictional sliding, and how that evolution can be externally influenced under external drivers are important questions, hard to investigate experimentally because the contacts themselves are generally difficult to access.
Here we focus on an elementary constituent of a general metal-metal interface, namely an ultrathin individual nanocontact, where recent rheological studies of crystalline gold nanocontacts [Nature 569, 393 (2019)] showed a dramatic and unexpected mechanical softening as a result of external oscillatory tensile stress.
The question which we address through realistic nonequilibrium molecular dynamics simulations is to what extent such mechanical softening  might influence the shearing habit of gold nanocontacts at room temperature. 
It is found that the shearing evolution, which occurs through a series of discrete slips, is indeed rheologically softened, even though not completely, by the oscillations.
Differences also emerge for different types of external oscillation, tensile or rotational.
The relevance of these results for future experiments will be discussed.
\end{abstract}

\section*{I-Introduction}

In the past two decades, studies of dry friction and wear \cite{Lantz2009}
have repeatedly shown how rapid load oscillations may cause a strong increase of lubricity between inert solid interfaces
that do not merge or cold weld \cite{Socoliuc.science.2006}
Theoretical arguments \cite{Ma.Urbakh.PRL.2015, Tshiprut.Urbakh.PRL.2005,Vanossi.Capozza.2012, Vanossi.2009.PRL,Guerra.pre.2008,Fajardo.prb.2014,Jeon.apl.2006}
essentially based on idealized models qualitatively suggested how, under these conditions, the externally imparted oscillations cause an easier disentanglement between facing asperities. 
%Here 
Conversely, here we are concerned with interfaces, such as those between ductile metals, where the local contact inevitably involves merging and/or cold welding, thus forming narrow solid bridges in correspondence with facing asperities, as discussed in classic literature \cite{Persson.Springer.2000, Persson.Tosatti.NATO.1996, Bowden.2001}.

At present, the detailed rheological evolution of such interfaces under shearing friction is insufficiently explored. For this situation, ``inert interface" type models %\cite{Ma.Urbakh.PRL.2015, Tshiprut.Urbakh.PRL.2005,Vanossi.Capozza.2012, Vanossi.2009.PRL} is 
are inadequate, and a new approach is called for. %Among other elements, some of the simulations aiming to describe such contacts \cite{He.2022} lack the important cohesion element leading to the intrinsic string tension of each nanocontact. \cite{Tosatti2001, Tosatti2005, Khosravi2022} % due to the nature of two different materials in contact, which results in an unstable frequency dependent simulation eventually describing an unstable states.

%As discussed in classic literature, \cite{Bowden-Tabot; Persson; BOOKS} %realistic 
%facing asperities that come into intimate contact at metal interfaces tend to merge locally, forming narrow metal bridges, often crystalline, between the two sides. 
Metal-metal nanocontacts, often crystalline, are reported in electron microscopy data, %most 
commonly in noble and near-noble metals, but not only \cite{Kurui2007, Kizuka2008, Ag_sato2012, Oshima2006}.
Experimental TEM images \cite{Carpick.Sato.2022, Ag_sato2012} in oscillation-free sliding metal contacts %also 
demonstrate a high degree of crystallinity of the bridging contact . {The spontaneous repair of mechanically induced stacking faults and twin boundaries, a repair predicted theoretically by Au nanocontact simulations \cite{Khosravi2022}, has most recently been reported experimentally in Ag \cite{Li.acsnano.2023}}.
%that is 
%{The local crystalline habit of many such nanocontacts persists despite the strong plastic deformation caused by nanocontact breaking and reforming during the sliding process.} 
The formation and presence of nanocontacts 
also determines
the electrical conductance of these metal contacts, whose overall Sharvin conductance might be roughly estimated as $G \sim N N_c G_0$ where $G_0 = 2e^2/h$ = 12.9 $\rm{(kOhm)}^{-1}$, for $N$ widely spaced nanocontacts in parallel, each of minimal cross section consisting of roughly $N_c$ atoms. % per unit area $A$.
%If each nanocontact has $\sim \nu$ atoms in its minimal cross section, that would roughly yield a total (Sharvin) ballistic conductance $G \sim N \nu G_0$ where $G_0 = 2e^2/h = 12.9 (kOhm)^{-1}$. 
Generally, nanocontacts interact with each other due to the combination of roughness and elasticity, as outlined in classic literature \cite{Persson.Springer.2000}.
% \AV{AV:paper 14 by Bo is more related to granular materials and powders}
% \Jin{Ok, fixed.}
When the interface is forced to slide under applied shear stress, the sparsely distributed nanocontacts are randomly sheared, 
%\AV{AV:comment:"sparsely distributed contacs" is fine! ...however I would not call the shearing of those "stochastic" (except for thermal effects).
broken, and reformed with mechanical friction and wear. \\

The shearing of single metal nanocontacts has already received attention in literature. Sato et al. \cite{Ag_sato2012,Carpick.Sato.2022} studied it experimentally in Ag.
More recent non-equilibrium molecular dynamics simulation by %Wang et al.
Yu et al. \cite{WangJin2020} remarkably rationalized these results. Additional TEM work described the shearing of W nanocontacts \cite{Wang.NatureNano.2022}. Studies showed that the shearing of a contact between two relatively large size asperities proceeds by %what one might call atomic 
stick-slip, %as expected for sliding between two crystal planes at the contact's middle cross section. The sliding evolved from 
e.g., atomic slips \cite{Wang.NatureNano.2022} at the beginning, evolving to multiple 
slips, close to breaking. 
%\ali{Maybe here we should clarify their 
However, the stick-slip is not always necessarily atomic stick slip. %Instead we can cite this newly found paper \cite{Wang.NatureNano.2022} for *atomic* stick-slip case.}
This evolution %was also 
is accompanied by a gradual weakening of the average shear stress.
%before breaking. 
%No %kind of 
The possible reduction of frictional shearing caused by additional oscillatory perturbations was generally not considered in these early studies.% ; the large contact cross sections studied would anyway have left little room for any rheological effects \AV{AV:True BUT... among the investigated tribological systems, those where tiny oscillatory perturbations are more effective are probably the ones related to SFA setups, which imply extended lateral contacts.}
%\ali{yes we have to clarify this}
%%%%%% for my info: surface forces apparatus (SFA)
%such as those 
We intend to pursue here such rheological effects in ultra-narrow metal nanocontacts, such as those recently studied experimentally\cite{Comtet2019}.  \\

% In the measured frictional shear stress, the underlying stochastic shearing/reforming nanocontact evolution would also reflect in the form of noise and fluctuations, both of mechanical and acoustical nature, and of electrical conductance.\\
%When frictional shear stress is measured, noise and fluctuations reflect the evolution of the underlying stochastic breaking and reforming nanocontacts. Noise thus contains much fundamental information; mechanical, acoustical, and electrical noise are unortunately rarely detailed and studied in experiments. \\ %either mechanical and acoustical in nature or of electrical conductance, is often considered as useless by-product and therefore discarded.\\
%One side curiosity will be to what extent the effects of imparted oscillations might or might not resemble those of temperature. As is known from simple models, the energy barriers that give rise to stick-slip friction are gradually overcome with a crossover to smooth sliding at high  temperatures - a regime appropriately dubbed \textit{thermolubric}.\cite{Krylov, Frenken} Room temperature, oscillation-free  shearing of a crystalline nanocontact does, as we shall show, proceed by sudden advancement-induced rearrangements akin to stick-slips. We will thus ask whether, by analogy, an additionally imparted mechanical oscillations might or might not similarly drive a crossover to a hypothetical {\it rheolubric} regime where sudden rearrangements and stick-slips gave way to smooth, liquid-like  shearing.\\
 
\begin{figure*}[!ht]
\centering
\includegraphics[width=\linewidth]{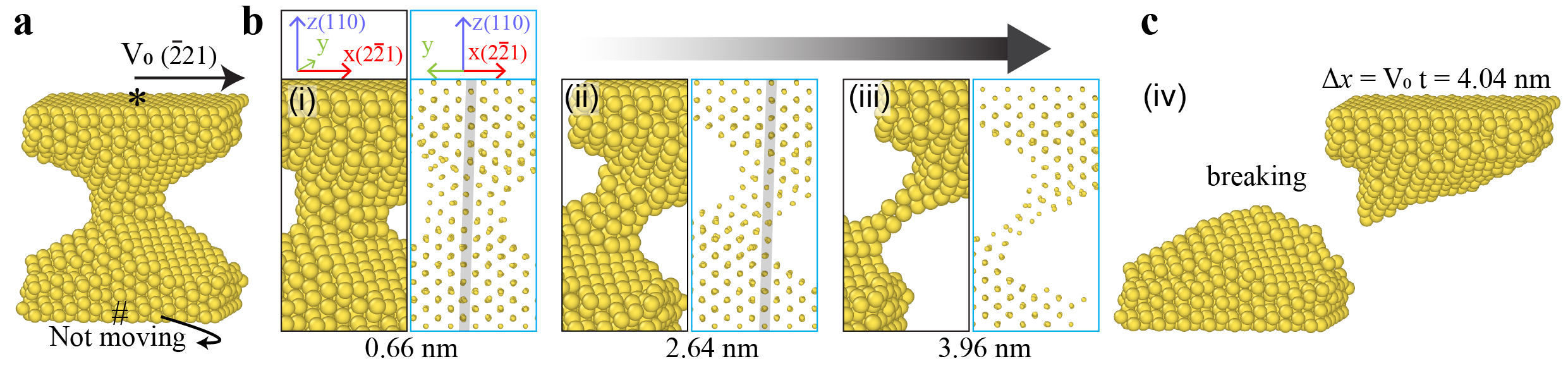}
\caption{
% \as{AS: name panels! } \ali{not necessarily}
Shearing along $(2 \bar{2} 1)$ of a Au nanocontact with constant velocity of
% \jin{$V_0=0.02$~m/s}
% \AV{AV: at room temperature?}
$V_0=0.02$~m/s at room temperature, no oscillations applied.
% \Jin{Two suggestions to the figure: (1) should we use $v_0$ rather than $\mathrm{V}_0$ in the figure? Also, maybe we should use italics for it. (2) the arrow on the top of the middle panel looks strange (and ``Time'' is also strange). How about put two smaller arrows on the bottom instead, between $\Delta x =0.66, 2.64, 3.96$~nm?}
% \Jin{note $\Delta x = V_0 t$ here.}
% \Ali{
Between $\Delta x=0.66$~nm and 2.64 nm the contact undergoes four stick-slip events, {each resulting in a narrower and longer nanocontact with one extra $111$ layer in it}. Two additional stick-slip events produce an umbilical braiding chain of gold atoms (4 atoms thick), owing to the inability to preserve crystallinity. The contact finally breaks after {shearing by } 4.04 nm.
% \as{The contact finally breaks after 4.04 nm}
Black and blue frames show the same moment of shearing %from two different perspective. 
{visualized differently}.
Detailed pictures of shearing with oscillations are provided in Supplementary S1.
% \Jin{Ali please add panel (a), (b)i, ii, iii, and (c) here.} \ali{Done}
}
\label{fig:1}
\end{figure*}

With that goal in mind, we simulated the shearing of a model gold nanocontact %,  where one anticipates that a
in presence of an applied mechanical oscillation. %should have an effect. 
Underlying and motivating this study is the recently discovered phenomenon of a seemingly liquid-like  collapse of mechanical %impedance 
stiffness as a result of strong strain oscillations in %gold 
such nanocontacts at room temperature\cite{Comtet2019}. That puzzling phenomenon was %very recently %explained 
subsequently 
shown \cite{Khosravi2022} not to involve melting, but rather to be due to a reversible stick-slip-like plastic evolution of the solid and crystalline  nanocontact, coupled with an ever-present string tension\cite{Tosatti2001,Tosatti2005}. Together, these %are the 
two elements %that 
lead to the observed negative nanocontact stiffness -- only apparently liquid-like -- %caused by oscillations in 
of a crystalline nanocontact under tensile oscillation.\cite{Khosravi2022}.\\

In this work, we first simulated the pure transverse shearing
of an ultra-thin gold nanocontact, where atomic stick-slip{-like shearing steps were} found at room temperature (see Fig.~\ref{fig:1}). 
Next, a vertical or rotational strain oscillation of increasing magnitude and fixed frequency was superimposed to the shearing displacement. %vertical or rotational strain oscillations of increasing magnitude (at fixed frequency). 
As we shall describe in Section II, the oscillation-softened nanocontacts experience some reduction of mean shearing friction, without actually losing the atomic stick-slip like steps. The nanocontacts in fact still maintain their solid crystal structure during all phases of shearing, despite the strong imparted oscillations. \\

Noise and fluctuations,  mechanical, acoustical and electrical,  also accompany %and reflect 
the shearing process. As shown in Section III, the analysis of that noise is particularly interesting and revealing for a nanocontact sheared in presence of  rotational oscillations. Section IV presents the dependence of shearing friction upon velocity and temperature, %which 
showing results in agreement with stick-slip expectations. Section V %will 
shows how the complex %dynamical 
mechanical 
response function (i.e., the conventional stiffness and damping) evolves during shearing. Our final conclusions, including prospectives for a proper ``rheolubric'' transition from stick-slip to smooth shearing at sufficiently low velocities and oscillation frequencies, close the paper in Section VI.  \\

%evolution of the underlying stochastic breaking and reforming nanocontacts. Noise thus contains much fundamental information; mechanical, acoustical, and electrical noise are unortunately rarely detailed and studied in experiments.  %either mechanical and acoustical in nature or of electrical conductance, is often considered as useless by-product and therefore discarded.\\
%Another 
%A broader question underlying this work is, to what extent the effects of imparted oscillations might or might not resemble those of temperature. As is known from simple models, the energy barriers that give rise to stick-slip friction are gradually overcome with a crossover to smooth sliding at high  temperatures - a regime appropriately dubbed \textit{thermolubric}.\cite{Krylov, Frenken} Room temperature, oscillation-free  shearing of a crystalline nanocontact does, as we shall show, proceed by sudden advancement-induced rearrangements akin to stick-slips. We will thus wonder whether, by analogy, an additionally imparted mechanical oscillations might or might not similarly drive a crossover to a hypothetical {\it rheolubric} regime where sudden rearrangements and stick-slips gave way to smooth, liquid-like  shearing.\\}

\section*{II-Nanocontact shearing simulations}

We conducted non-equilibrium molecular dynamics (NEMD) simulations including a standard Langevin thermostating at room temperature. The simulation setup included approximately 2500 atoms (Fig. \ref{fig:1}).
The upper and lower gold leads were composed of two rigidly stacked FCC(110) lattice planes. Four thermostated planes, each consisting of 576 atoms, were attached to the leads on each side, connecting them with an initial column-like nanocontact shape  of approximately 2 nm in length. 
With the present choice of crystal orientation, the column is made up of first-neighbour (110) atomic chains, a crystallographic choice
that yields a nanocontact particularly robust and resistant to thinning and breaking  %\AV{AV:It would be nice to show, at least in SI, a sketch of this initial setup configuration}
%\ali{Sure -- Jin please take care of this}.
After careful relaxation and annealing, surface atoms migrated and the column turned into the more realistic long-lived shape of Fig. \ref{fig:1}. The narrowest transverse cross section of the relaxed nanocontact comprised $N_c \sim $ 26 atoms. Electrically, that would correspond to about $N_c$ ballistic channels. 
Many such ultrathin nanocontacts could of course be similarly generated and studied. As shown in previous work \cite{Khosravi2022}, their rheological behaviour is however essentially equivalent. We argue therefore that studying just one should suffice to address their generic shearing behaviour. \\

To minimize the 
%possible % AS: is this really a possility? I take it for granted that thermostatting the neck  would influence the dynamics
influence of Langevin damping on the subsequent NEMD dynamics, the thermostat was only applied to 
three atomic layers of mobile atoms outside (i.e. above and below, see Fig 1)  %well outside 
the %are physically relevant 
nanojunction region. %\as{AS: can this region be defined more quantiatively?} \AV{AV: ...or at least shown in the sketch of the initial configuration}.
Additionally we verified that, within a significant range of values of the Langevin damping parameter \cite{Benassi.PhysRevB.2010}, 
% \AV{AV: please cite here: A. Benassi, A. Vanossi, G.E. Santoro, E. Tosatti, Phys. Rev. B 82, 081401(R) (2010).}, 
the rheological %response 
{behaviour} of the system was essentially independent of our specific chosen value. The whole setup was 
fully equilibrated %thermally and mechanically relaxed \AV{is it clear what we mean here by 'thermally and mechanically'?} 
before the shearing simulations. Shearing was performed by moving the upper lead with a constant velocity $V_0=0.02$~m/s along  the $(2 \bar{2} 1)$ direction (our $x$-axis), orthogonal to the %initially 
vertical (110) {initial} nanocontact axis. Shearing was also carried out at increasing velocities, up to $0.2$~m/s. When it was accompanied by oscillations (to be introduced below), either vertical or rotational, the frequency of oscillations was generally set at $1$ GHz.
These large velocities and frequencies, imposed 
by current computational limits, are of course way higher than their typical experimental counterparts \cite{RevModPhys.2013}.
% \AV{I would cite here our Rev. Mod. Phys. 85, 529 (2013).}
However, as our previous studies \cite{Khosravi2022} 
showed,  %that 
the room temperature rheological behaviour of gold nanocontacts %at room temperature r
{is expected to remain} the same across many orders of magnitude of these physical parameters. In particular, a loss of the so-called necking-bellying jumps in favour of a smooth rheological evolution of a nanocontact of the size studied here is not expected until a very low oscillation %crossover 
frequency of order 600 Hz. A further point worth stressing here is that none of the rheological manoeuvres realized in %the present 
{our type of } simulations\cite{Khosravi2022}, and {\it a fortiori} in experiments,\cite{Comtet2019} implies a rise of temperature. All heat produced by shearing and by oscillations is effectively removed to the thermal bath {(experimentally embodied by both leads), } which stabilizes room temperature. Therefore no change of friction should or could be attributable to {Joule heating causing}  a hypothetical onset of ``thermolubricity''\cite{Krylov-Frenken2014}. 
In that respect, it should also be noted that the  heat conductivity of gold {or other metals} is more than an order of magnitude larger than that of our atomistic simulations, which do not include electronic effects. %\as{AS: worth mentioning that also in experiments heat is efficiently carried away by electrons?}

% It has been shown before that the general behaviour of these nanojunction is the same above a crossover frequency around $600$Hz up to $10$ GHz \cite{Khosravi2022}.

%One of the difficulties to link between experiment and simulation is the vibrational frequencies used in experiments which usually are in the range of tens of KHz, but in simulations are 3 to 6 orders of magnitude larger. This issue can be addressed by expecting that the two regimes of frequency will reflect the same physical principles. . It was fortunate that the essential physical characteristic, stick-slip, could only be transformed into viscous sliding due to thermal fluctuations at frequencies below 600 Hz \cite{Khosravi2022}, well below the experimental ones \cite{Comtet2019}. So we confidently remain in high frequency regime, for simplicity of calculation.

\begin{figure}[!h]
\centering
\includegraphics[width=1\linewidth]{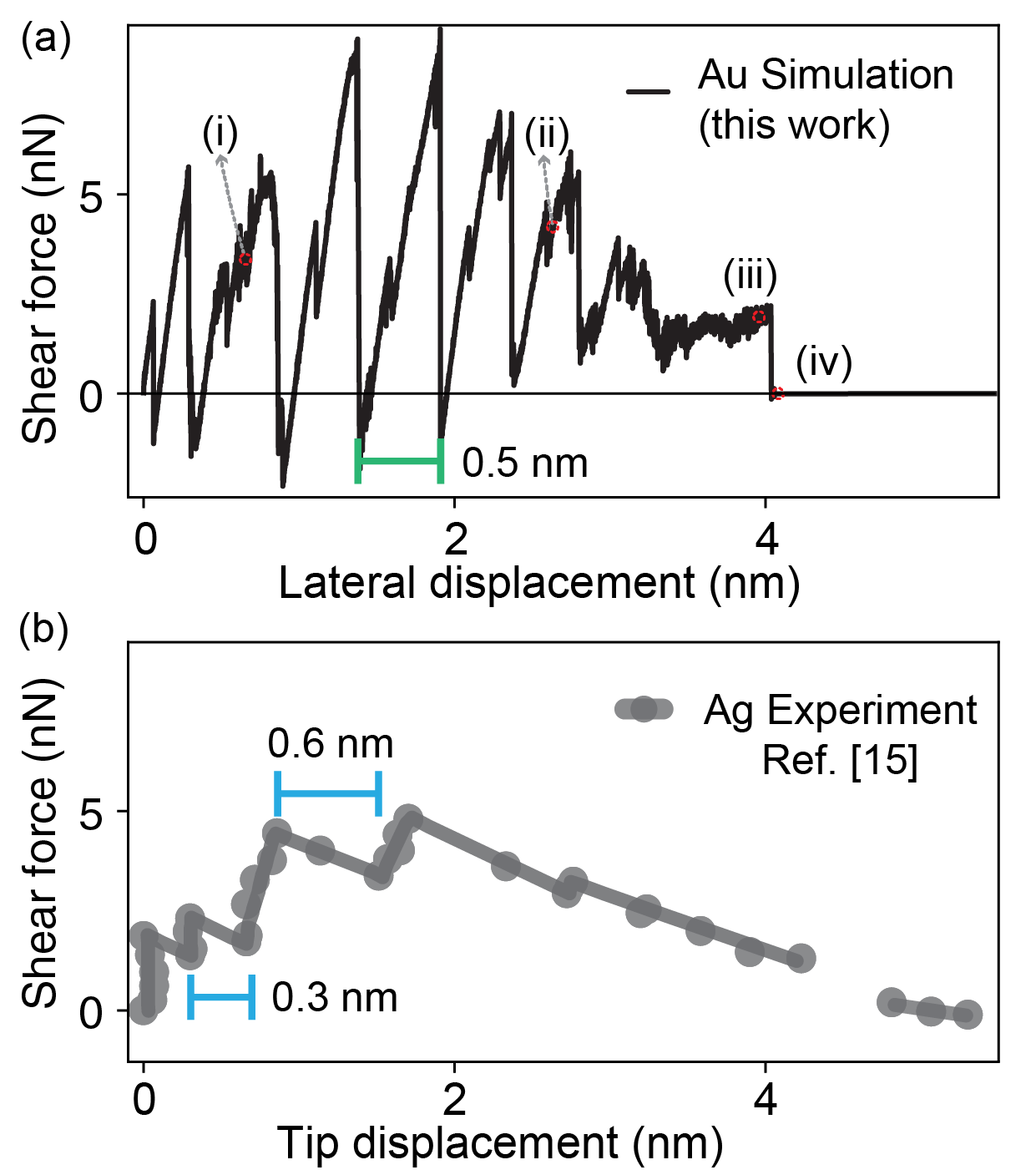}
\caption{{(a) Instantaneous shear force versus shear distance of the simulated Au nanocontact of Fig. \ref{fig:1}. Force rises during sticking, and drops at slips where shearing advances. (b)  Experimental shear force of Ag nanocontact, from Ref. \cite{Ag_sato2012}. In both cases (a) and (b) there is atomic stick-slip. The slip is close to $2R$ in Ag, ($R$ is the atomic radius) and to $2R\sqrt{3}$ in Au, where shearing is along $(2 \bar{2} 1)$.  Note the similar shear
force magnitudes, despite the  $\sim$ 5  times larger size of the experimental Ag contact. That size difference, together with a larger damping, is probably responsible for the multiple atomic slips and the smaller force jumps of Ag slips. Indices (i) to (iv) in the figure correspond to the three states during sliding in Fig. \ref{fig:1}(b).}}
% \ali{put a b c, and refer to fig 1}
% \Jin{I think the main difference between the two curves is the definition of displacement: the black is the displacement of the lead and the grey is the displacement of the tip.}
% \Erio{Jin, the only difference I see is that the experimetal tip has a finite stiffness, the theoretical lead has infinite stiffness...Is that what you mean?}
% \Jin{Yes, the infinite stiffness is the difference. So, I do not understand why we say: ``a feature lost in a thicker junction'' here in the caption.
% I suggest to write something in agreement with the main text: ``a feature lost in the system with smaller stiffness''.} %\as{AS+AV: we need to discuss this sentence (and the text). Also in Ag they have atomic stick slip (as they say in the paper).}
\label{fig:2}
\end{figure}

\subsubsection*{The starting point: shearing without oscillations}

\begin{figure*}[!ht]
\centering
\includegraphics[width=\linewidth]{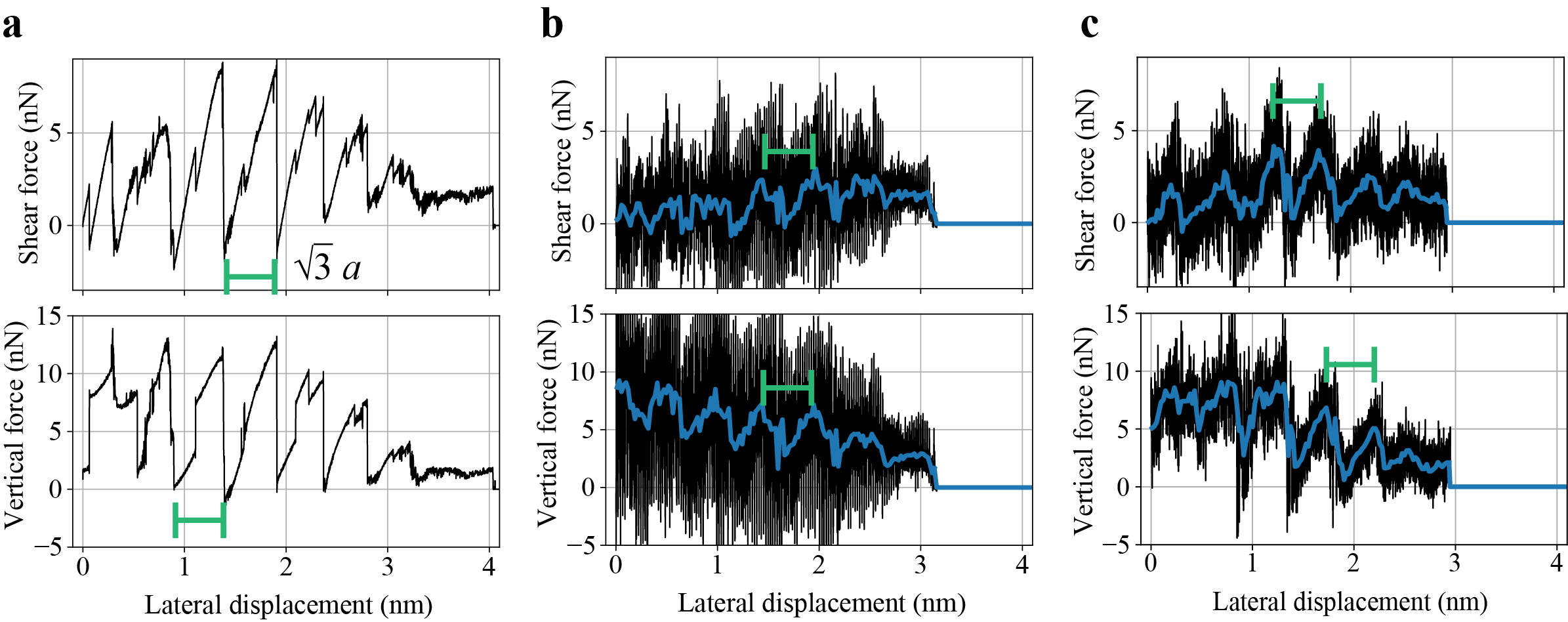}
\caption{Horizontal and vertical force {evloution in} simulated shearing of the Au nanocontact of Fig.1.
(a) No oscillations; (b) Large vertical oscillation ($a_0 = 0.22$ nm);  
(c) Large twist oscillation ($\theta_0 = 30$ degrees).
Blue lines are averages drawn for better visibility. Green bars show the length of slips, {whose} value $\approx 0.5$ nm reflects the Au lattice spacing  $\sqrt{3} a$ in the $2\Bar{2}1$ sliding direction.
% \Jin{I suggest to explain what is $a_0$ here in the caption. And it might be better to mark it in the figure.}
% \as{AS: why only the first raw has labels? Please, add also d,e,f. Also I suggest to add on top of each column the driver: Pure shear, shear+vertical oscillation, shear+twisting}
% \ali{make the x axis, take away the extra 1 2 3 4 things +  mark $a_0$ in the figure}
% \Jin{Fixed.}
}
\label{fig:3}
\end{figure*}

In the first of set of shearing simulations, % conducted without oscillations, 
we applied a lateral %shearing with 
constant velocity {$V_0 = 0.02$}~m/s to the upper lead, without oscillations.
The shear force evolution between the two leads shown in Fig. \ref{fig:2} (black) is read  off the simulation, showing the sawtooth profile typical of stick-slip advancing steps.
%, through which the nanocontact shearing takes place. AS: redundant
As the stick-slip-mediated shearing proceeded, the nanocontact gradually deformed into a narrower and narrower oblique junction, thinning down and eventually breaking apart, see Fig. \ref{fig:1}.
% f. 
Note that the size of these stick-slip-like jumps 
%size , with constant magnitude close to  0.5 nm, is shown by the green bar in the force signal Fig. \ref{fig:3}\textbf{a}.\\
is constant (approximately 0.5 nm, see green bar in Fig.~\ref{fig:3}a).\\

Structural examination, detailed in 
% Supplementary $S????$,
Fig. \ref{fig:1}
showed that each slip is due to a sudden shear-induced elongation. The nanocontact rearranges by thinning down while adding to its length, at each slip, one extra (111) plane, an actual solid slice. Shearing slips are therefore %entirely 
akin to the oscillation-induced necking jumps earlier reported with vertical oscillations and no shear.\cite{Khosravi2022}. 
That being ascertained, the partially sheared nanocontact becomes increasingly %slanted, 
oblique at first sight by a continuous amount as the shearing proceeds. Thus the question arises, if the nanojunction obliqueness grows continuosly, why do shearing slips strictly retain the atomic step length prescribed by the two leads, as if the nanocontact remained vertical? 
\\

Although not visible in Fig. \ref{fig:1}, the answer is structural 
% (see Supplementary S ???),
(see Fig. \ref{fig:8}a)
and quite interesting. Oblique and slanted as it is, the nanocontact {\it always retains unbroken  vertical (110) atomic chains that connect the two leads}. At the each slip, the sudden necking rearrangement, while increasing by one the number of (111) cross section planes, stops precisely once at the earliest distance where unbroken (110) vertical atomic chains can reform. And that is %precisely  
exactly 
the $(2 \bar{2} 1)$ lattice spacing in both leads, that is $2R\sqrt{3}$,  where $R = 0.148 $ nm is gold's atomic radius.  The rationale behind this unsuspected shear slip quantization therefore appears to reflects atomic jumps between successive adhesion energy minima, which the surviving unbroken (110) chains realize between the two leads after each slip.  \\

It is instructive to compare this force profile with the experimental one: the gray line in Fig. \ref{fig:2} reports the force measured in Ref.~\cite{Ag_sato2012} for a  thicker silver nanocontact.
The experimental shearing force also exhibits stick-slip steps. It has an overall non-monotonic magnitude from initially small to maximum after a few slips, to a final decline before breaking. That behaviour had already been described and understood by simulations \cite{WangJin2020}.  We superpose it here to our own nanocontact shearing simulation, so as to highlight similarities and differences. 
Despite the smaller cross section
% \as{and the different chemistry} 
of our nanocontact, the  overall similarity is remarkable. %Yet, the shearing slips of the Ag contact are not quantized and atomistic, probably due to the much larger size and the finite mechanical stiffness of the experimental setup (our simulation corresponding to infinite stiffness) .  % including atomic stick-slips and force magnitudes \ali{Maybe I'm wrong, but I think in Ag experiment there is no atomic stick-slip, just plastic yielding}. \\
%, despite the two different metals, Au and Ag. \\

There are, on the other hand, important differences.
Our thin nanocontact shear slips are all atomic and,  as explained above, correspond to a single step in the $(2\bar{2}1)$  direction. That differs from the {shearing} evolution %from single to multiple slips 
of the thicker Ag contact, where %no quantization is detectable. 
{quantizaton of slip length, directly detectable in Fig.2a, occurs first in units of nearest-neighbor distance $2R$, but successively evolves to multiple steps.}
Another difference is the magnitude of force jumps, which is much larger in our simulated system. 
This very likely reflects vastly different relaxation times and  mechanical stiffnesses of experimental and simulation shearing setups.
One advantage of the stiff simulation setup we adopted is to emphasize direct access to {\it noise}, which is paramount in stick-slip shearing.
The shearing noise in our simulation was strong, both in the (horizontal) $x$-directed shearing force and in the (vertical) $z$-directed force, see Fig. \ref{fig:3}\textbf{a}.
%That is actually quite natural, for more than one reason. % AS: too colloquial, in my opinion
This is a reasonable result for at least two reasons. 
First, the vertical string tension, established and steady during the sticking period, must undergo a  jump at each slip, while returning to a steadier, slightly smaller value at next sticking, when unbroken (110) chains reform, of course less numerous. 
Another is the increasing tilt angle between the nanocontact body and the vertical, causing a growing tilt %decreasing level of parallelism 
of the slipping layers away from the $x$-direction of shearing. %\as{AS: unclear sentence... what is a "decreasing level of parallelism"? }. \\

%{A second difference is that the experimental slip step lengths, determined as they are by the unspecified Ag nanocontact geometry, also change in the course of the shearing. A likely explanation for that could be that the initial single atomic slips gradually evolve into multiple slips. }

\subsubsection*{Shearing with vertical { (tensile)} oscillations}
%\as{AS: this paragraph is quite unclear to me. Are we talking of pure oscillatory or there is also shearing? At the beginning it seems so, then there's a "prior to shearing", then we talk about shearing again. The paragraph ends with "there are the two elements" and yet I cannot understand what these two elements are...}
In this %the {subsequent} 
main set of simulations we %next 
studied the nanocontact shearing while also applying an oscillatory $z$-displacement $A(t)= a_0 \sin{\omega t}$, symmetrically to both leads. Increasing oscillation magnitudes $a_0$ were explored, 
%reaching values 
{up to} $a_0 \sim 0.22$ nm, 
%where prior to shearing  the nanocontact stiffness rheologically turned from positive to negative.\cite{Khosravi2022}. 
at which the nanocontact stiffness turns from positive to negative \cite{Khosravi2022}.
The center of the nanojunction 
%zone 
underwent an oscillating tensile/compressive deformation. 
For small amplitudes, 
%that simply
this
added an oscillating noise on top of the pure shear signal discussed previously (see supplementary section S2). 
At 
%higher
larger
oscillation amplitudes $a_0 > 0.22 $ nm, deformations are known to develop necking and bellying jumps, which repeat reversibly in each cycle of oscillation during shearing. 
In all conditions including this, a large average tensile force (an intrinsic string tension \cite{Tosatti2001,Tosatti2005}), spontaneously persists beween the leads. Thanks to it, even a solid nanocontact with a crystalline core  resists breaking at necking and reversibly heals at bellying.\cite{Khosravi2022}
%The intrinsic string tension \cite{Tosatti2001,Tosatti2005}, % with 
%{(thanks to}
%which even a solid nanocontact resists breaking at necking and %favours healing 
%{heals}
%at bellying), maintains in this condition a large average tensile force. 
%These are the two elements that %drive 
%{This reversible plastic deformation}
%the rheological softening of the contact prior to shearing. 
This mechanism remains at work during shearing,  %while
%, in the course of time, 
during which however 
the nanocontact thins down 
%while maintaining its stick-slip advancing behaviour.\\
via stick-slip advances.

Fig. \ref{fig:3}\textbf{b} shows the extreme case of $a_0 = 0.22 $ nm vertical oscillation amplitude. As Supplementary Movie S1 shows, the overall effect of oscillatory necking/bellying \cite{Khosravi2022} is to make the nanocontact cross section narrower after each slip. 
%It
The nanocontact 
survived in spite of this narrowing, continuing to exert its string tension between the two leads. 
%That
This
kind of "live" behavior makes a ductile metal contact 
%really different
qualitatively different
from rigid unreactive systems and models. %\as{AS: if the term of comparison is "ductile metal contact", I'd use "systems" rather than "models".}.
Despite %such 
% a violent and % AS: the subject here is amplitude, not oscillation. How can an amplitude be violent?
the large oscillation amplitude at frequency of 1 GHz, stick-slip shearing advances still persisted (blue line in Fig. \ref{fig:3}\textbf{b}), their characteristic length (green bar), same as %the previous case of shearing 
without oscillation.

\begin{figure*}[!ht]
\centering
\includegraphics[width=\linewidth]{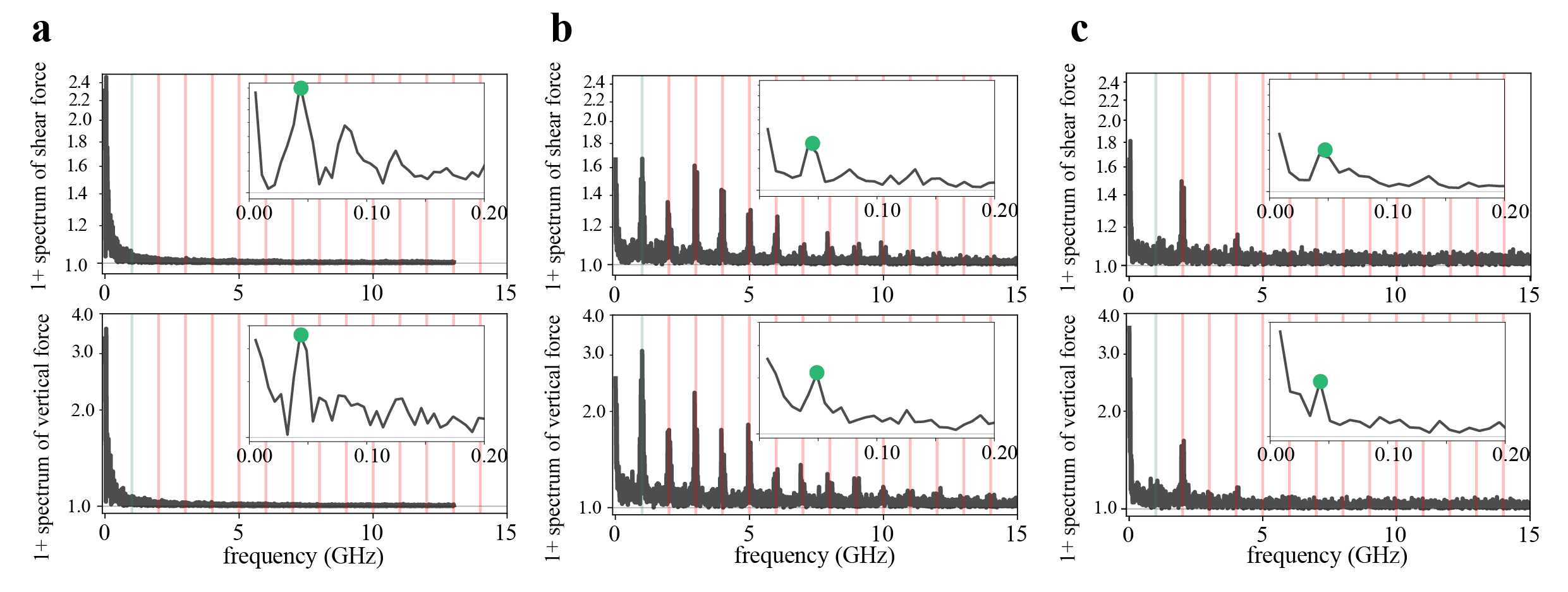}
\caption{%Spectrum of 
{Force Fourier %transform 
spectrum} associated with Fig. \ref{fig:3} for (a) non-oscillatory shearing contact; (b){ shearing under large vertical oscillations } %extreme vertical oscillation 
($a_0 = 0.22 nm$); %while shearing, (
(c) { shearing under large rotational oscillations 
($\theta_0 = 30^{\circ}$)}. % degrees. 
For better visibility, the y-axis is scaled logarithmic  and the spectrum is shifted by +1. 
The inset is a zoom-in {of the low frequency range} %interval. 
The green dot shows the stick-slip peak at $\approx$ 42 MHz, %signaling the length-scale 
{corresponding to the green bar periods} in Fig. \ref{fig:3}. Note %that 
{how the high frequency integer peaks are modulated by the  low} stick-slip frequency --see Supplementary Section S3.}\label{fig:4}
\end{figure*}

\subsubsection*{Shearing with rotational ("twist") oscillations}

In a subsequent set of simulations, instead of vertical oscillations, we rotated one lead relative to the other with an oscillatory twist of the form 
%$\theta(t)= \theta_0 \sin{\omega t}$
$\theta(t)= \theta_0 \sin{\omega t} + \Phi$
%\as{AS: this is the only place $\theta(t)$ is presented, include the offset $\Phi$ and then say it's 0 in the first paragraph.}
($\Phi$ = 0 corresponding to fully aligned leads) while at the same time actuating the shearing without any other perturbations. Because the stick-slip advancements reported in the two previous subsections were connected with interplanar 2D lattice slidings, a reasonable expectation was that oscillatory twists might deeply alter the shearing mode.

The force profile of shearing with a  torsional oscillation as large as $\theta_0 = 30 ^\circ$ and $\Phi$ = 0 is shown in Fig. \ref{fig:3}\textbf{c}. 
The shearing slips and noise are quite similar to those with tensile oscillations. %is visibly smoother \as{AS: Not that clear to me}. 
The nanocontact breaks earlier, confirming that twist oscillations disturb the shearing more effectively.% than oscillating strains.  
Yet, even in this %supposedly 
more disruptive case, the average frictional force is not %\as{considerably} 
visibly smaller. Importantly, and to some extent surprisingly, the stick-slip shearing 
%habit
behaviour also
persisted with the same atomic step length (green bar) as in the two previous cases.
%In order to minimize possible artifacts, we performed additional 
This scenario persisted in shearing simulations with $\Phi \neq$ 0, also discussed in Section V.
%\ali{not in supp. anymore. Now in Sec?}
Despite differences of details and noise, the stick-slip shearing habit was found to persist even in that case. Only by pushing $\theta_0$ to extreme values as large as 60$^\circ$ was it possible to cross over to a rheologically smooth shearing 
%habit
behaviour.
. 
%In the next Sections we shall gain more understanding on these diverse data with more quantitative tools.

\subsection*{Vertical force and string tension} 

In all shearing conditions, the stick-slip jumps of the friction force are systematically accompanied by even stronger jumps of the  lead-lead  normalforce.  An important feature of vertical forces that the nanocontact transmits, between the two leads ( Fig. \ref{fig:3}), is its systematically positive average value. That reflects the natural  string tension, which a ductile metal nanocontact will display in all finite temperature conditions. It can be rationalized as follows. In a hypothetical liquid junction, the total free energy, minimized by reduction of the surface area, gives rise to capillary attraction between the two partner bodies. A solid crystalline nanocontact does not possess bulk capillarity, yet the mere chemical potential difference felt by an atom between the junction (an unfavoured site) and a bulk solid lead (a more favourable site) has consequences. The first is that atoms will slowly but inevitably drift thermally from the junction to the leads, provoking thinning and eventual breaking. The second is to create a thermodynamically based string tension between the leads qualitatively similar to that of a liquid neck, despite the inner crystallinity of the nanocontact \cite{Tosatti2001,Tosatti2005}. At every horizontal slip, the vertical force jumps %in phase with 
with a certain phase with respect to the horizontal shearing force. Generally, the nanocontact  behaves, as an oblique active spring. %The magnitude of vertical force jumps is almost a factor 2 larger than the shear counterparts. 
The vertical force average -- the underlying string tension -- gradually drops as the shearing proceeds, clearly due to progressive thinning of the nanocontact prior to breaking, shown by Fig \ref{fig:1}.\\

The vertical force traces of Fig. \ref{fig:3} complements the simultaneous information on the evolution under the oscillatory perturbations by the horizontal force on the shearing process.
The first element is the persistence of vertical jumps approximately { 90$^{\circ}$ out of} phase with the horizontal shear slips, and  nearly twice as large. 
As the shearing proceeds, the progressive drop of the vertical force average 
%with tensile oscillations, and even more with twist oscillations, is stronger than without oscillations.  
in the presence of vertical oscillation is stronger than the pure shearing case, see Fig.~\ref{fig:3}a,b. This is even more evident in the case of twisting oscillations, see Fig.~\ref{fig:3}c.

\begin{figure*}[!ht]
\centering
\includegraphics[width=\linewidth]{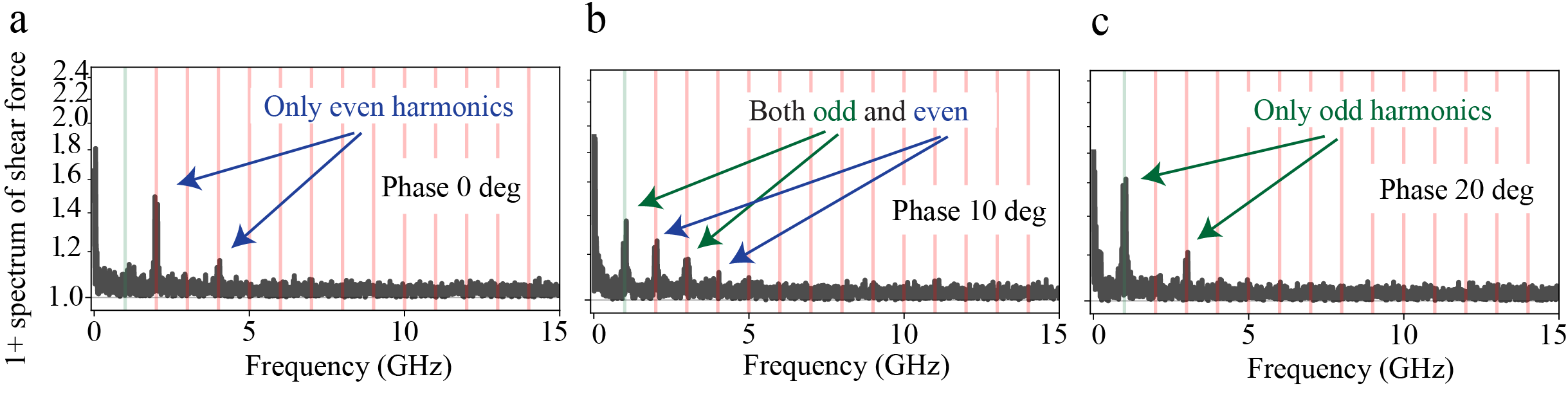}
\caption{Fourier analysis of the force noise during shearing accompanied by a rotational oscillation ($ \theta_0 =30^\circ$), for different crystal orientation, phase $\Phi$ of the two leads . 
%Left, 
(a) $\Phi$ =0, perfect alignment of the upper and lower leads (same as in Fig. \ref{fig:4}c), where only even harmonics appear. 
%Middle,  
(b) Intermediate misalignment $\Phi$= 10$^\circ$, where both odd and even harmonics are nonzero.
%Right,
(c) Large misalignment $\Phi$= 20$^\circ$, where only odd harmonics appear.
% \jin{X-label's unit can be $\omega_0$ rather than GHz.}
% \ali{Updated, chronological order}
}
\label{fig:5}
\end{figure*}

\section*{III-Shearing noise analysis}

The previous Section presented the global picture of the rheological effect of oscillations on the shearing habit of a thin nanocontact. More quantitative insights can be obtained by Fourier analysing both shear (horizontal) and vertical force traces of Fig. \ref{fig:3}. That also permits addressing the large mechanical noise which accompanies the shearing.\\

Fig. \ref{fig:4} shows the Fourier spectrum of the force traces of Fig. \ref{fig:3}. The initial shearing without oscillations %\jin
{(Fig.~\ref{fig:4}a)} presents a single peak around $\approx 42$~MHz (marked with green dot in the inset).
That is the washboard frequency of atomic stick-slips
%\jin
{$f_w = V_0/\lambda$, where $V_0 = 0.022$~m/s}
is the simulated sliding velocity and $\lambda \approx 0.5$~nm  is the atomic stick-slip length along $2\Bar{2}1$ (marked with the green bar in Fig. \ref{fig:3}). In the shearing with oscillations, either vertical or rotational 
{(Fig.~\ref{fig:4}b,c)}, a weaker washboard frequency peak survives, %if both oscillation frequency and the sliding velocity are above the viscous regime - $\approx$ 600 Hz \& 300 $nm/s$-. This is true 
even in extremely large oscillation amplitudes.
Therefore the oscillations do not cause true rheolubricity in the shearing of the nanocontact.  This weakening marks nevertheless a rheological softening of the shearing habit. That is due to the rapid oscillation-induced reversible plastic deformations, taking place while the slower shearing takes place. These plastic deformations were found to consist of necking-bellying for vertical oscillations \cite{Khosravi2022}, while they consist of interplanar angular locking-unlocking for rotational oscillations.\\
In both of these two cases -- see Fig.~\ref{fig:4}\textbf{b,c} -- a high frequency noise appears in the form of higher harmonics of the imparted oscillation (1 GHz in the case shown). The sinusoidal imposed oscillation elicits a strong non-linear response. Qualitatively, we may associate the total number of peaks showing up to the number and intensity of rearrangements that the nanocontact needs to undertake in order to preserve its crystalline structure after each stick-slip events.\\

These nonlinear peaks in the noise offer potential insights into structural changes.
With rotational oscillations in particular, different combinations of n-th harmonics, such as odd or even, appear and disappear based on the relative crystalline orientation of the two leads. 
As shown in Fig. \ref{fig:4}\textbf{c}, when upper and lower leads have the same crystal orientation,  shear force Fourier peaks possess only even harmonics. That is because the shearing slip generally takes place while the nanocontact is maximally twisted, that is when the twist magnitude is $\theta_0 + \Phi$.
%\Jin{The range of twist angle should be given here, otherwise ``maximally twisted'' is unclear.}
The nanocontact spends the longest time near that top dead point, where the originally straight (110) chains of gold atoms are subjected 
%by twist to a very costly "braiding".
forced by the imposed twist into a energy costly ``braiding''.
For vanishing average 
%static %AS: what is static here?
twist $\Phi=0$, there the two top dead points per rotational oscillation cycle, $\omega t $ = 0 and $\pi$,  are two equivalent.
The shear force time dependence can in that case be crudely assimilated to
\begin{equation} 
 p(t) = |\; \sin(\omega_0  t ) \;|
\end{equation}
As a result, Fourier shear force peaks just at twice of fundamental frequency $\omega_0$, even harmonics. \\
% The shear force time dependence can be crudely assimilated to

% \ali{After today's discussion 18th Oct, I'll check how to retrieve this formula in such a way that only even will succeed. }
% \begin{equation} 
%  p(t) = \Sigma_{n}{ \Theta{[\omega_0 t -  \pi(n+1/4) ]} - \Theta{[\omega_0 t -  \pi(n+3/4) ]} } ,
% \end{equation}

% {where $\Theta{[x]}$ is Heaviside's function. The Fourier transform of Eq.(1) has peaks precisely at $2n \omega_0$.}\\

% \erio{PLEASE CHECK!}
% \ali{Dear Erio: I've checked, the above formulation results in all 2*odd peaks, 2*even will be zero. Apparently it is not possible to have all even, with such formulations. I just learned this phenomena is called in signal community as "**Duty cycle**". I think we can also use their language. Suggestion are in green color. }\\

% \erio{I am confused Ali. Please look at https://mathworld.wolfram.com/FourierSeriesSquareWave.html  which seems to contradict what you say... }\\
% \ali{This function also is all odd. all even components are zero. no?}

The overall misalignment $\Phi$ about the vertical axis of the upper lead crystal orientation with respect to the lower one, additional to the oscillatory one $\theta_0 \sin{\omega t} $ brings an interesting change. %\as{AS: what is the amplitude $\theta_0$ here?} 
Increasing misalignments $\Phi= 10^\circ$ and $20^\circ$ caused a clear evolution of the spectral noise peaks %in the noise spectral, 
as shown in Fig. \ref{fig:5}.
With $\Phi$= 10$^\circ$, both odd and even harmonics turned nonzero. By increasing to  $\Phi$= 20$^\circ$, even harmonics disappeared and only odd harmonics remained. \\

This is the spectral distribution expected for  50\% duty cycle, such as %with a time dependence of the shear force closer to a "rectified" form 
% \ali{maybe add the rec. pictures?}

\begin{equation} 
%\begin{split}
    % p(t) =\sum_{n} &\Theta{[\omega_0 t -\pi(2n +1/4)]}\\
    % -&\Theta{[\omega_0 t - \pi(2n + 5/4)  ]}  , 
    p(t) = \Theta[ {\sin(\omega_t)}],
%\end{split}
\end{equation}  % ALI: I CHANGED FROM 3/4 TO 5/4 TO HAVE ALL ODD
% \ali{ This formulation also works: $p(t) = \text{sign}[\; \sin(\omega_0  t ) \;] $}. 

% \Ali{where $\Theta{[x]}$ is Heaviside's step function \as{AS: where's the Heaviside?}. \ali{We took it away}
(where $\Theta$ is Heaviside's function) whose Fourier transform of Eq.(1) has peaks precisely at $(2n+1) \omega_0$.
The underlying physics is that forward shear slips now take place only when twist reaches $\theta_0 + \Phi$, but not when it reaches the other dead point $\theta_0 - \Phi$, which is much smaller.
% {whose Fourier transform contains only odd powers.}\\
% \erio{PLEASE CHECK!}\\
%Other ratio of duty cycles, directly reflects in Fourier transform in various forms, are independent of the choice of fundamental frequency of oscillation itself.}
For more general parameters $\theta_0 $ and $\pm \Phi$ of  the rotational oscillations, the inequivalent efficiency of the two dead points directly reflects in the even-odd peak distribution of the noise Fourier transform. A distribution which, we find,  is relatively independent of the oscillation frequency $\omega_0$ .
These results might be further developed for possible applications such as discovering the crystal orientations of nanocontacts, a feature usually only visible by TEM.
%\Ali{ for unburied setups.}
\\

%{Try $\Phi$ = 30  degrees  (that is, +-15)? Should be even stronger odd-only behaviour!!! }\\
%\ali{We tried, but it seems both peaks appear in that case. }

The present analysis is limited to mechanical noise, but the underlying structural phenomena suggest extrapolating mechanical noise to electrical current noise as well. As expected, and also shown by recent experiments \cite{Latha.Fu.2021}, tensile force and conductance jump simultaneously at each of the successive contact thinning events that precede breaking. That is clearly explained by the drop in the nanocontact minimal cross section. The separate smooth evolution of force and current between the jumps are less correlated, with  conductance as stable as the cross section, and pulling force building up to reach the thinning point. %  more they have different behaviour in between breaking jumps that calls for further experiments and explanation.

%{{\it Vertical force} An important feature of vertical forces that the nanocontact transmits between the two leads ( Fig. \ref{fig:3}), is a systematically positive average value. That is just the natural {\it string tension} which a ductile metal nanocontact will display in all finite temperature conditions. It can be rationalized as follows. In a hypothetical  liquid junction, the total free energy, minimized by reduction of the surface area, gives rise to capillary attraction between the two partner bodies. A solid crystalline nanocontact does not possess bulk capillarity, yet the mere chemical potential difference felt by an atom between the junction (an unfavored site)  and a bulk solid lead (a favourable site) has consequnces. The first is that atoms will slowly but inevitably drift thermally from the junction to the leads, provoking thinning and eventual breaking. The second is to create a thermodynamically based string tension between the leads qualitatively similar to that of a liquid neck, despite the inner crystallinity of the nanocontact. \cite{Tosatti2001,Tosatti2005} At every horizontal slip, the vertical force jumps in phase with it, the nanocontact  behaving as a stretched oblique spring. Its average -- the underlying string tension -- gradually drops as the shearing proceeds. That reflects the progressive thinning of the nanocontact prior to breaking, shown by Fig 1. }

\begin{figure*}[!ht]
\centering
\includegraphics[width=\linewidth]{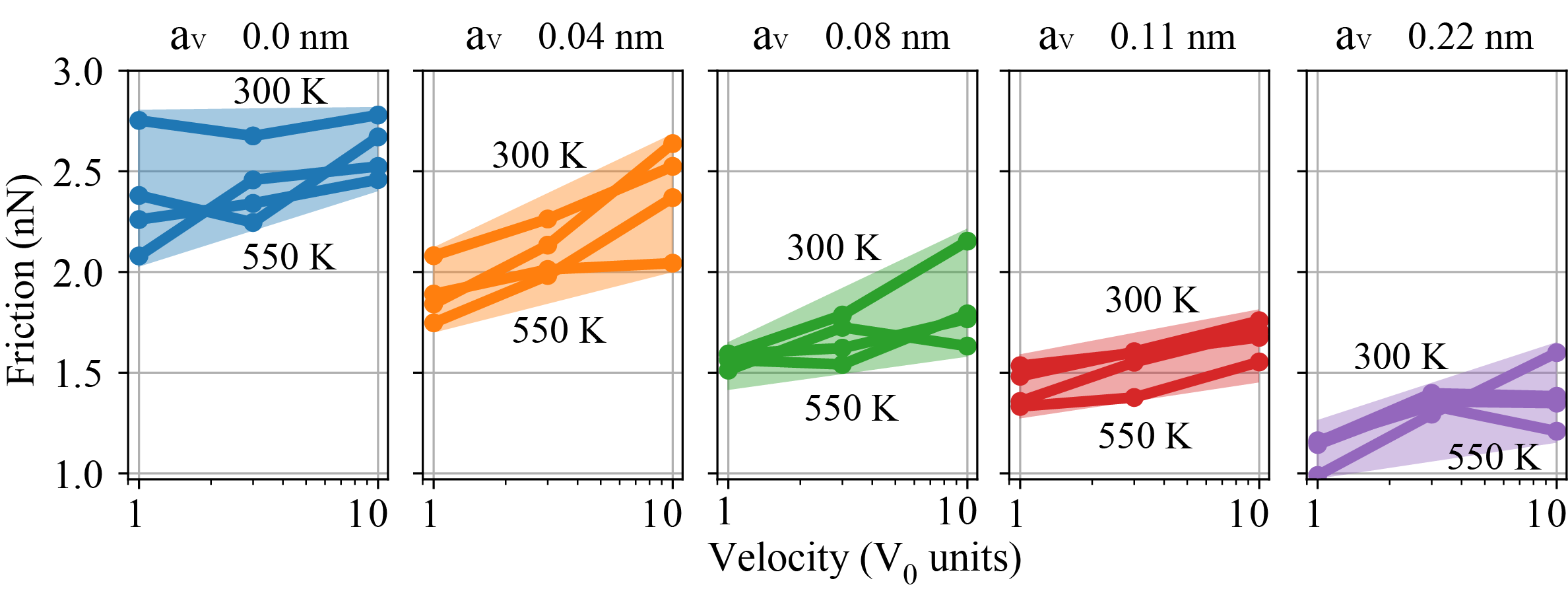}
\caption{Shear friction force as a function of velocity $v_0$. Our lowest velocity (same in all cases in the main text) was $V_0=0.022$ m/s. Each line refers to a given temperature $T$ and vertical oscillation amplitude $a_0$.{ Note the clearly sublinear rise with velocity}.
% \Jin{X-label should be velocity rather than Log velocity. Fixed}
%{ ERIO: WHAT ABOUT TWIST OSCILLATIONS?
%\ali{Unfortunately, I didn't run this analysis for the TWIST case}
}\label{fig:6}
%From left to right, shearing is accompanied with zero to large of vertical oscillation amplitude, $a_v$. }\label{fig:6}
\end{figure*}

\section*{IV-Velocity and temperature dependence}

We can now return to our initial motivation. Namely, discovering to what extent external mechanical oscillations, tensile or angular, could transform
%, without extra help by temperature, % AS: seems redundant and breaks the flow
the shearing 
%habit 
behaviour
of a thin nanocontact from stick-slip to viscous -- a process we tentatively branded  "rheolubricity". The velocity dependence of frictional shear, expected to be logarithmic for stick-slip and linear for viscous sliding, can provide the most direct diagnostic \cite{Vanossi.RevModPhys.2013}. To explore and verify this velocity  hallmark, we ran additional simulations at a variety of speeds and temperatures. \\

The average frictional shearing stress, conventionally defined as the midpoint average of horizontal force traces such as those of Fig. \ref{fig:2} and \ref{fig:3}, was extracted and plotted versus velocity in  Fig. \ref{fig:6} for various vertical oscillation amplitudes and for increasing temperatures.  Each line represents the velocity dependence of friction at a given temperature and vertical oscillation amplitude $a_v$. The simulation thermostat always applied to the leads and not to the nanocontact) ensures in all cases  that the mechanical heat introduced by oscillations and shear is completely
completely %effectively}
conducted away. These further simulations were too limited in number to be statistically accurate; the resulting uncertainty is represented as a shaded area comprising upper and lower temperatures used. While that spread could not permit detailing a delicate logarithmic velocity dependence, the overall behaviour is clearly
%and undoubtedly much% AS: a bit too strong 
weaker than linear, in agreement with the persistence of stick-slip already reported in the previous sections.  Results for twist oscillations, (not explored at this stage) are also expected to be similar.
%ERIO: IS THIS ACTUALLY THE CASE, ALI??
%\ali{Yes, sure we expect it to be even more weaker than this. But I don't have proof or simulations}
In conclusion, coherently with the previous sections, the mechanical oscillations do bring a %certain 
noticeable
amount of rheological softening of frictional shearing, although without proper rheolubricity at the frequencies considered. As will be mentioned in Conclusions, that outcome should
change at sufficiently 
low oscillation frequencies, where adiabatic behaviour is expected. These low frequencies (estimated to be below 600 Hz for the present nanocontact) fall outside of the range accessible to direct simulation. \\
%In addition, these simulations clearly show that our thermostat is effective enough that even at large temperatures, e.g. 550K, large dissipated heat during oscillation has been conducted away so that the velocity dependence is still logarithmic. 

\begin{figure*}[ht!]
\centering
\includegraphics[width=0.9\linewidth]{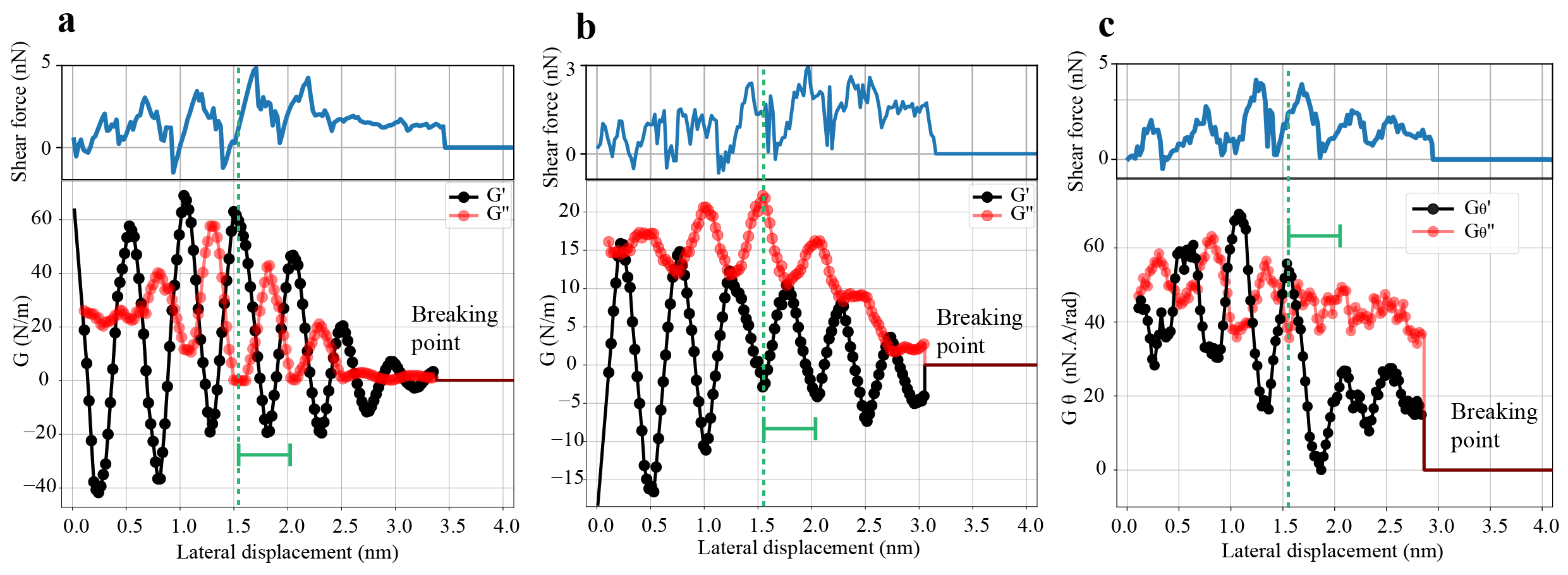}
\caption{
Evolution of the complex dynamic vertical 
response $G$  for shearing under vertical oscillation with amplitude (a)  $a_0= 0.08$~nm (b) $a_0= 0.22$~nm, and (c) of the complex dynamic rotational response $G_\theta$ for shearing under rotational oscillation $\theta_0 =30^\circ$.
Green marks indicate the periodicity of slip, same as in Fig. \ref{fig:3}, and \ref{fig:4}. 
The green dashed line matches the {extrema of the dynamical module near the middle of each stick-slip. The nanocontact is essentially elastic after each slip. 
For reference, the blue curves show the corresponding shear force as a function of displacement. 
% The phase shift $\Phi$  relative to the external vertical oscillation (dashed sinusoidal line) evolves from in phase to out of phase when the amplitude goes from small to large. Ali: I'm sorry for causing this confusion, the phase shift is between the two G' in (a) and (b)
}
% {
% ERIO: COULD WE ALSO ADD IN THE TOP PANEL A DASHED LINE WITH THE EXTERNAL SINUSOIDAL OSCILLATION SUPERPOSED TO THE BLUE SHEAR LINE? ONLY FROM THAT ONE CAN REALLY SEE THE PHASE $\Phi$ AND UNDERSTAND WHAT HAPPENS...
% }
% \ali{but Erio, that is 180 cycle of oscillation. I think nothing will be visible. Maybe I have to clarify this phase $\pi$ is meant relative to stik-slip periodicity}
}
\label{fig:7}
\end{figure*}

\section*{V- Dynamical response function}

% In the course of shearing,  
% a nanocontacts undergoes a structural evolution that rheologically affects its mechanical properties. Our imposed oscillatory perturbation, $p(t)=p_0 \sin{\omega t}$, elicits dynamically a certain response. 

% In general not sinusoidal, but with a large sinusoidal component $r(t)= r_0 \sin{(\omega t +\phi)}$. For a vertical oscillation, $p_0 =  a_0$, and $r_0 =  f_0$, the latter is the best fit to the sinusoidal part of the vertical force intensity. 
% The ratio $G = G' +i G''= (f_0/a_0) \exp(i\phi)$ is the complex dynamical response function, whose real and imaginary parts are technically referred to as storage module (more commonly "stiffness"), and loss module. 

% In the rotational case, $p_0 =  \theta_0$, while $r_0 =  \tau_0$, the latter is the sinusoidal best fit to the torque intensity, and the corresponding response function $G_{\theta}=G'_{\theta} +i G''_{\theta}= (\tau_0/\theta_0) \exp(i\phi)$. 
% These are the complex dynamical response
% whose evolution we wish to explore during the nanocontact shearing. 
% \\

% Start first with shearing in presence of the vertical oscillation. There is first of all a permanent static string tension force $f_{st}$ pulling the leads together. 
% Then, the oscillatory vertical force of our interest is extremely noisy -- see Fig. ????. That required averaging over a sufficient number of cycles -- 10 in our case. 

In the course of shearing,  
a nanocontacts undergoes a structural evolution that rheologically affects its mechanical properties. An imposed oscillatory perturbation $p(t)=p_0 \sin{\omega t}$, elicits dynamically a certain response, 
in general not sinusoidal. Although 
%not really  a
formally outside the realm of
linear response, the (generally large) sinusoidal force component $r(t)= r_0 \sin{(\omega t +\phi)}$
% \cite{Comtet2019} as a mechanical impedance -- a dynamical response function. 
% \Ali{as a fitting on experimental\cite{Comtet2019} or simulation data can be used to calculate mechanical impedance -- a dynamical response function. }
{can be used to calculate mechanical impedance \cite{Comtet2019} -- a dynamical response function. }

For a vertical oscillation, $p_0 =  a_0$, and $r_0 =  f_0$, the latter` is the best fit to the sinusoidal part of the vertical force intensity. Note the response of interest (Fig. \ref{fig:3}) is very noisy at room temperature which requires averaging over several cycles in order to get a reliable fit (in our case, 10).
The ratio 
$G = G' +i G''= (f_0/a_0) \exp(i\phi)$ 
% $G = G' +i G''= (f_0/a_0) \exp(i\phi)$ % AS: to be coherent with the next one % Ali: in that case I would have need to update the figure, sorry maybe after first round of review..
is the complex dynamical response function, whose real and imaginary parts are technically referred to as storage module (more commonly "stiffness"), and loss module. 
In the rotational case, $p_0 =  \theta_0$, while $r_0 =  \tau_0$, the latter is the sinusoidal best fit to the torque intensity, and the corresponding response function is $G_{\theta}=G'_{\theta} +i G''_{\theta}= (\tau_0/\theta_0) \exp(i\phi)$. 
These are the complex dynamical response
whose evolution we wish to explore during the nanocontact shearing. \\

\begin{figure}[ht!]
\centering
\includegraphics[width=0.8\linewidth]{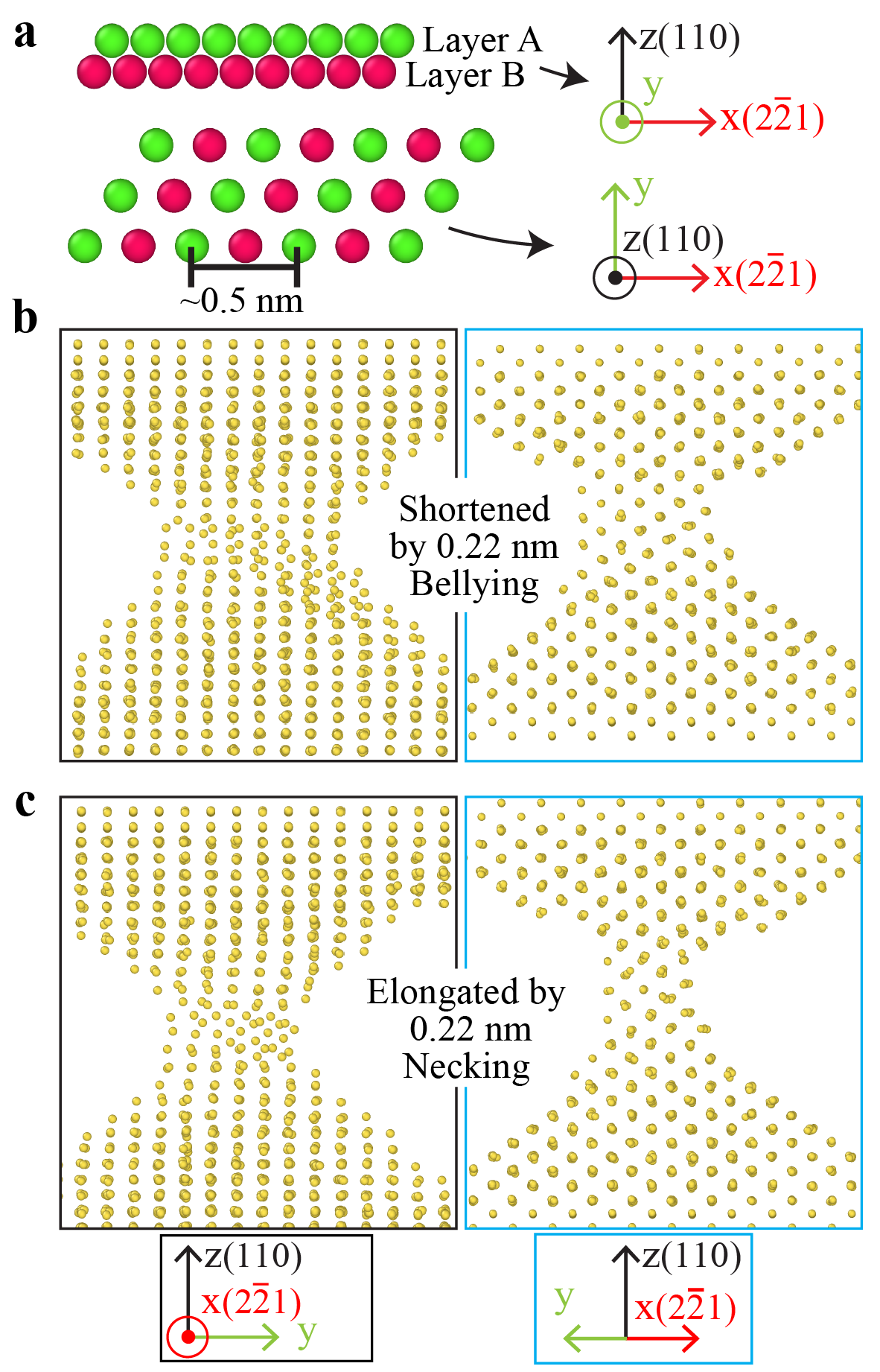}
\caption{
(a) The lattice structure, green and red represent A and B layer in the $110$ direction. In the case of negative stiffness, A prefers to match with B and B with A, an alternation.
(b) and (c) are showing the twinning moment occurring reversibly at each cycle of vertical oscillation near the dead-points where the contact spend must of its time. 
This only occurs when the vertical oscillation amplitude is large enough (here $0.22$ nm) giving rise to negative stiffness. \cite{Khosravi2022}
This particular layer arrangement shifts the stick-slip points by half the lead's surface lattice spacing $(1/2) 2R \sqrt{3}$ = 0.25 nm, thus %altering solution of AB matching BA, 
creating a local AB $\to$ BA stacking fault, while maintaining the verticality of 110 chains. Panels (b) bellying, and (c) necking moments from two different angle.
}\label{fig:8}
\end{figure}

Results are shown in Fig. \ref{fig:7}. 
All modules undergo dramatic up-down evolution in the course of shearing.
Opposite extrema of stiffness $G'$ and loss module approximately coincide with the mid point of the shear force slips -- the green dashed line.
For the rotational oscillation, stiffness remains always positive, independent of amplitude $\theta_0$, and its undulation has maximum stiffness almost at matching lattice points along $2\Bar{2}1$ direction of AB layers $\approx 0.5$ nm.
For vertical oscillation, the result strongly depends on the amplitude $a_0$. 
For small $a_0=0.08$ nm (Fig. \ref{fig:7}a), the stiffness 
$G'$ (initially positive) is in phase with the 
stick-slip shearing as it was without oscillation, %or with rotational oscillation -- 
see the force profile in Fig. \ref{fig:3} or Fig. \ref{fig:7}a,c.
For large amplitude $a_0=0.22$ nm, the stiffness 
$G'$, initially $<0$ owing to the large tensile oscillation \cite{Comtet2019, Khosravi2022}) 
successively takes positive and negative values as successive slips carry out the shearing, with a phase reversal of  $\pi$, relative to
the small amplitude case with initial $G' >0$ . 
A close look into the lattice structure of Fig. \ref{fig:8}a 
reveals the origin of this shift.
The phase shift is due to a structural change in the nanocontact induced by the large vertical oscillation. At this large oscillation amplitude, the staggered AB stacking of mid-junction crystal planes, with $2\Bar{2}1$ spacing of 0.5 nm, is unstable and locally deforms into {BA, with a shift of 0.25nm in lattice spacing.} Thus the running of $G' $ from negative to positive and back with a phase shift of $\pi$ relative to the standard case of Fig. \ref{fig:8}a also represents a structural diagnostic for this rheologically created local stacking fault --  an interesting consequence of negative stiffness.
\footnote{
To further explore our understanding of negative stiffness$G'$, extra simulations are performed with no shearing but two simultaneous oscillation along vertical and orthogonal direction ($y$). See Supplementary Section S4.
}

\section*{VI- Discussion and conclusions}
% \ali{I like this section, well written}

%Underlying this work is the broad question: to what extent the {rheological friction softening caused by} imparted oscillations %on friction 
%might or might not resemble those of temperature? \as{AS: this was not mentioned clearly before this point...}. As is known from simple models, temperature has a %strong 
%{weakening} effect on dry friction. In particular, the energy barriers that give rise to stick-slip can be thermally overcome with a crossover to smooth sliding at high  temperatures - a regime appropriately dubbed \textit{thermolubric}\cite{Frenken2005}. 
The oscillation-free shearing of a crystalline nanocontact  does, as we have shown, proceed by sudden advancement-induced rearrangements akin to stick-slips. It is thus natural to wonder whether, by analogy to thermolubricity (the crossover from stick-slip to smooth sliding caused by themperature \cite{Frenken2005}), an additionally imparted mechanical oscillations might similarly drive a crossover to a hypothetical {\it rheolubric} regime where sudden rearrangements and stick-slips could give way to smooth, liquid-like  shearing. We have presented arguments backed by realistic simulations showing that mechanically imparted high frequency oscillations, either vertical, or rotational, do reduce shear friction but do not eliminate stick-slip. The nanocontact rheological softening reduces the frictional shear but at least at frequencies above the KHz it does not suffice to produce rheolubricity, that would change the nature of the shearing process from stick-slip to fully smooth and viscous. 
%That % AS: does this refer to the concept just discussed? Then "this" is better than "that", no?
This failure goes hand in hand with the known persistence of a crystalline structure in the oscillated nanocontact which opposes a smooth shearing shearing. Crystallinity is unexpectedly robust also thanks to the pervasive string tension present in the nanocontact in all conditions.\\ 

The persistent stick-slip nature of shearing entrains a number of consequences. First, and unmistakable, the velocity dependence of friction is much weaker than linear as would be expected for viscous sliding.  Our simulated velocity dependence is indeed compatible with the logarithmic dependence typical of stick-slip. 
Second, there is a strong mechanical shear noise, with nonlinear mixing between the washboard and high harmonics of the fundamental oscillation 
%frequencies
frequency. As an interesting side note, the predicted Fourier spectrum of shear noise with rotational oscillations may 
%change from only even to only odd 
present only even or only odd
overtones depending on the relative crystalline orientation of the two leads. That is an unanticipated result of potential value for structural diagnostic. Accompanying all the above, during the stick-slip shearing the nanocontact complex dynamical modulus -- stiffness and loss module -- jumps dramatically at each slip. 
As a result of  planar alternation, their phase shifts by $\pi$ in the case of negative stiffness,  experimentally an observable element.
\\

A relevant question to be addressed before closing is the domain of validity of these results, based as they are on simulations with relatively large shearing velocities and extremely large oscillation frequencies. What if these velocities and frequencies were much lower? Thermodynamics actually demands that shearing at sufficiently low velocity should necessarily reach the thermolubric regime, where stick-slip must disappear\cite{Frenken2005, WangJin.2023.Colloquium}. Similarly, oscillating a nanocontact at a sufficiently low frequency should wash out the necking-bellying phenomena, with crossover to a smooth, nearly adiabatic evolution \cite{Khosravi2022}. Simulations cannot approach these low frequency and velocity regimes, but  extrapolations are nonetheless possible as follows. Concerning frequency, %it was estimated \cite{Khosravi2022}
%in the past that 
the crossover oscillation frequency for a nanocontact such as that studied here it was estimated \cite{Khosravi2022}  around 600 Hz. Therefore, even for very modest oscillation frequencies, from tens of kHz to MHz, the stick-slip scenario just presented should still hold -- of course with numerical modifications in case of different nanocontact sizes. Similarly, by equating the washboard frequency $\sim V_0/a$ to the crossover frequency (600 Hz in our case) one can estimate a low velocity limit  $\approx 300$ nm/s, where stick-slip shearing should turn thermolubric and  viscous.  
Even in that case the nature of rheological softening remains different from that of thermolubricity.\\

The above estimates suggest a speculative experimental proposal in conclusion of this work where rheolubricity was sought and not found within the parameter ranges of our simulations. Suppose a nanocontact shearing experiment was initially started at a velocity just above the crossover velocity, without oscillations. The shearing would be stick-slip and noisy, as described.  If subsequently 
%at that point 
the oscillations were turned on, the regime should change, with a transition from stick-slip to viscous shearing -- both friction and noise should drop. The same result could be obtained at constant velocity by switching oscillations from above to below the crossover frequency. Rheolubricity might therefore be attainable in future experiments.  \\

\section*{Acknowledgments}
The author acknowledge support from ERC ULTRADISS Contract No. 834402. Support by the Italian Ministry of University and Research through PRIN UTFROM N. 20178PZCB5 is also acknowledged.

\bibliographystyle{unsrt}
\bibliography{ref}

% \clearpage
\onecolumn

\section*{Supplementary information}

\subsection*{S1. The evolution of breaking junction}
\begin{figure*}[h!t]
\centering
\includegraphics[width=\linewidth]{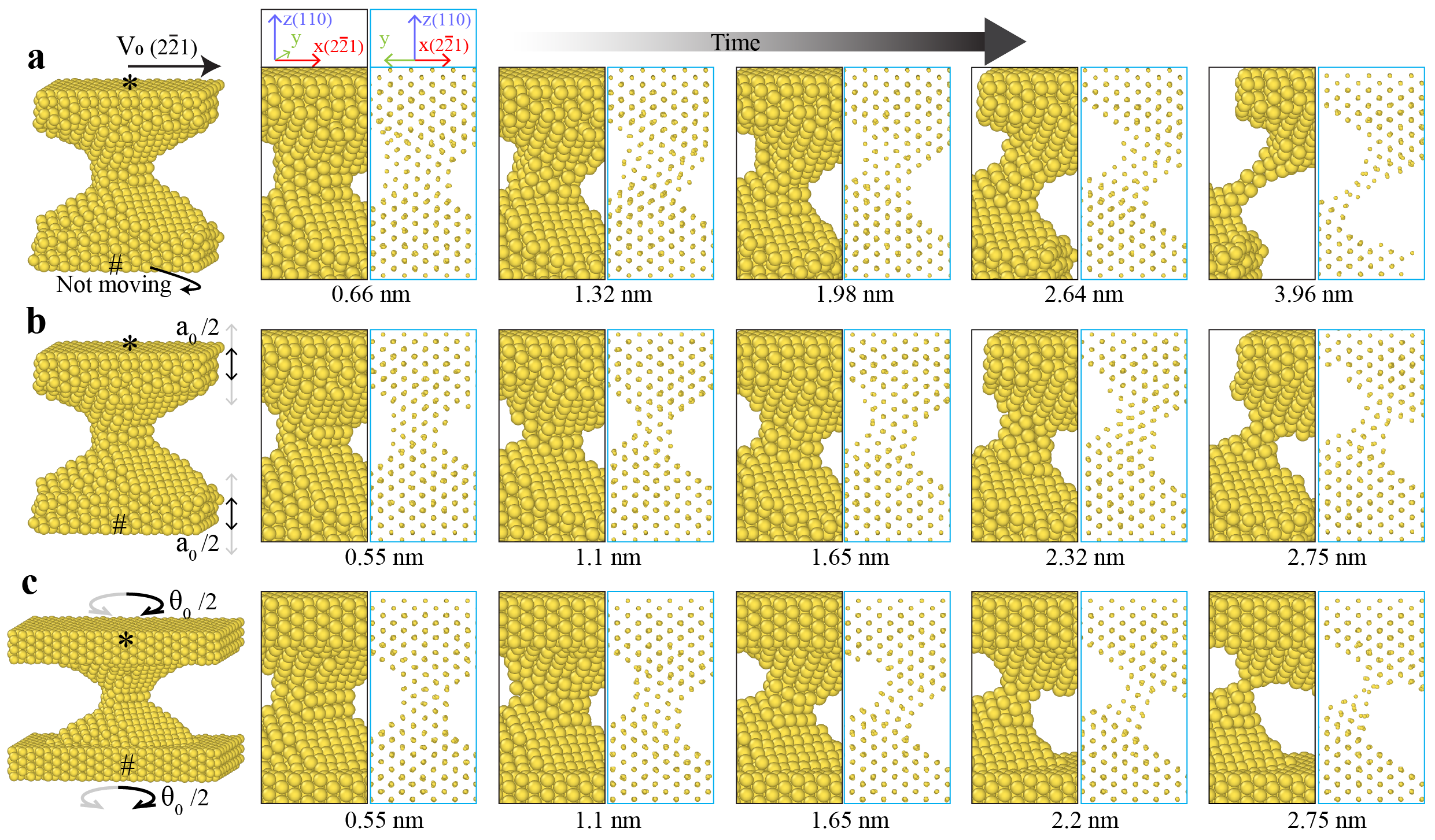}
\renewcommand\thefigure{S1}
\caption{Lattice structure evolution. (a) While shearing without oscillation, the same as identical to Fig. \ref{fig:1}; (b) with vertical oscillation $a_0$=0.22 nm ; and (c) with rotational oscillation $\theta_0$ = 30 degrees. 
Black and blue frames show the same moment of shearing from two different perspective. 
Shearing velocity $V_0$ = 0.02 m/s; frequency of oscillation in both vertical and rotational case set is 1 GHz.
% \ali{put the vertical shaded thing}
}\label{fig:S1}
\end{figure*}

\subsection*{S2. Small vertical oscillations}

\begin{figure}[!ht]
\centering
\includegraphics[width=\linewidth]{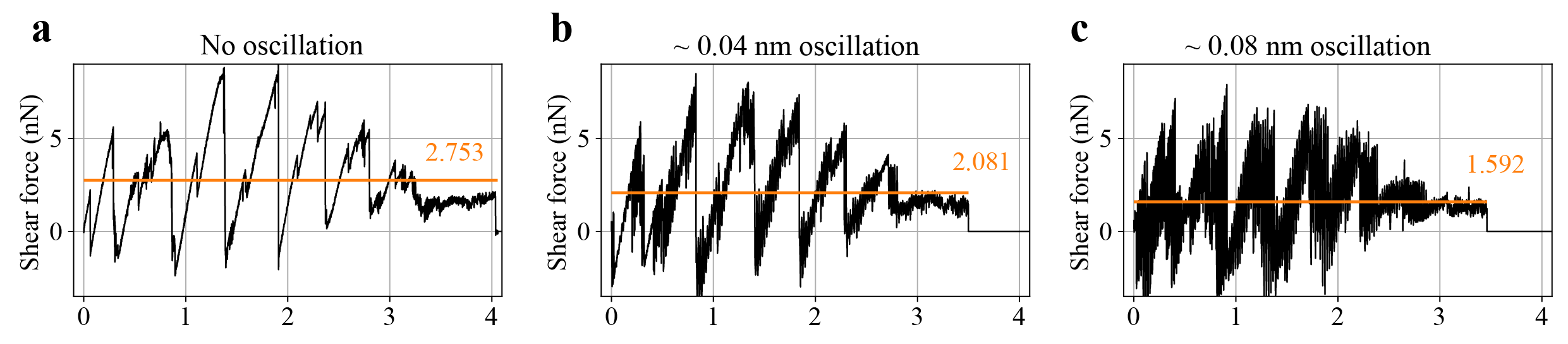}
\renewcommand\thefigure{S2}
\caption{Shear force signal for (a) shearing without oscillation (b) with vertical oscillation of small amplitudes 0.04 nm (c) and 0.08 nm}\label{fig:S2}
\end{figure}

In Fig. \ref{fig:S2} we show how the effect of small oscillations, $0.04$ nm and $0.08$ nm is building up a noise in the original signal of shearing without vertical oscillation. This rheological noise is reducing the average friction overall -- see the orange line.

% \newpage
\subsection*{S3. Coupling of washboard and oscillation spectra}
\begin{figure}[!ht]
\centering
\includegraphics[width=\linewidth]{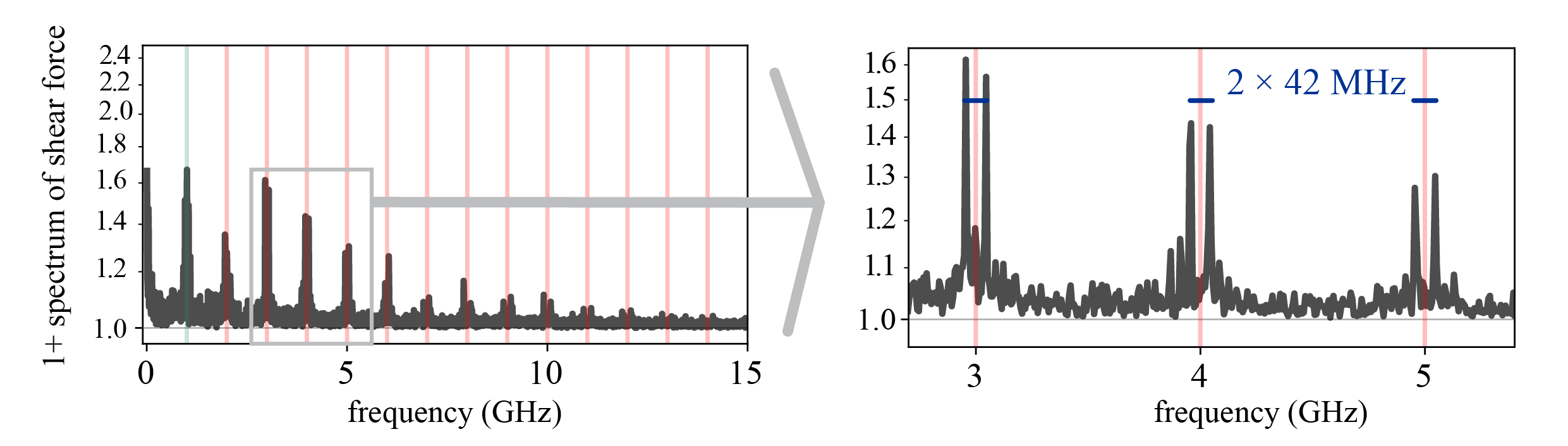}
\renewcommand\thefigure{S3}
\caption{Frequency analysis of shearing force accompanied with large vertical oscillation of $a_V = 0.22 nm$, the same case as in Fig. \ref{fig:4}b. The zoom-up in the right, shows a hidden overcoupling feature, with separation of twice the stick-slip frequency, $\approx 42$ MHz. This feature should disappear in experiment, without eliminating the rearrangement peaks.   }\label{fig:S3}
\end{figure}

%We can observe an overcoupling in our signal analysis. 
It is interesting to analyse the fine frequency structure of our shearing force spectra. As the inset in Fig. \ref{fig:4} shows, the washboard frequency is that the of stick-slip peaks, $\approx 42$ MHz. The external oscillation  peaks are at  the higher frequency of 1 GHz. Owing to the strong nonlinearity of the system, and to the high shearing velocity,  $0.022 m/s$ in our simulations, the two phenomena couple and interfere, and the two frequencies mix. As a result, all oscillation-related peaks are split by  the much smaller washboard frequency, as shown in Fig. \ref{fig:S3}. %just before and after each excited high frequency rearrangement peaks. This coupling is a result of the high shearing velocity of $0.022 m/s$ in our simulations. 
In experiments with much lower sliding velocities, the coupling should be reduced. If moreover much lower oscillation frequencies could be used, the nanocontact rhelogical behaviour could evolve from necking/bellying to adiabatic, whereby all high frequency peaks would disappear altogether. 
% An order of magnitude estimate given for the crossover frequency in the present nanocontact is
Based on an order of magnitude calculation, the crossover frequency of the current nanocontact is estimated to be $600$ Hz.

\subsection*{S4. Evolution of vertical $G$ with lateral oscillation}

This time, we run a series of simulations of the same contact oscillated laterally ($y$ direction) and vertically ($z$), but with no shearing.
Results confirm our understanding of negative $G'$, turning its sign once it's accompanied with large lateral oscillation. 
% \as{AS: is this linked to a structural transition in the nanocontact?} \ali{yes, it is addressed in the G sections. To be written more clearly..}

\begin{figure}[!ht]
\centering
\includegraphics[width=\linewidth]{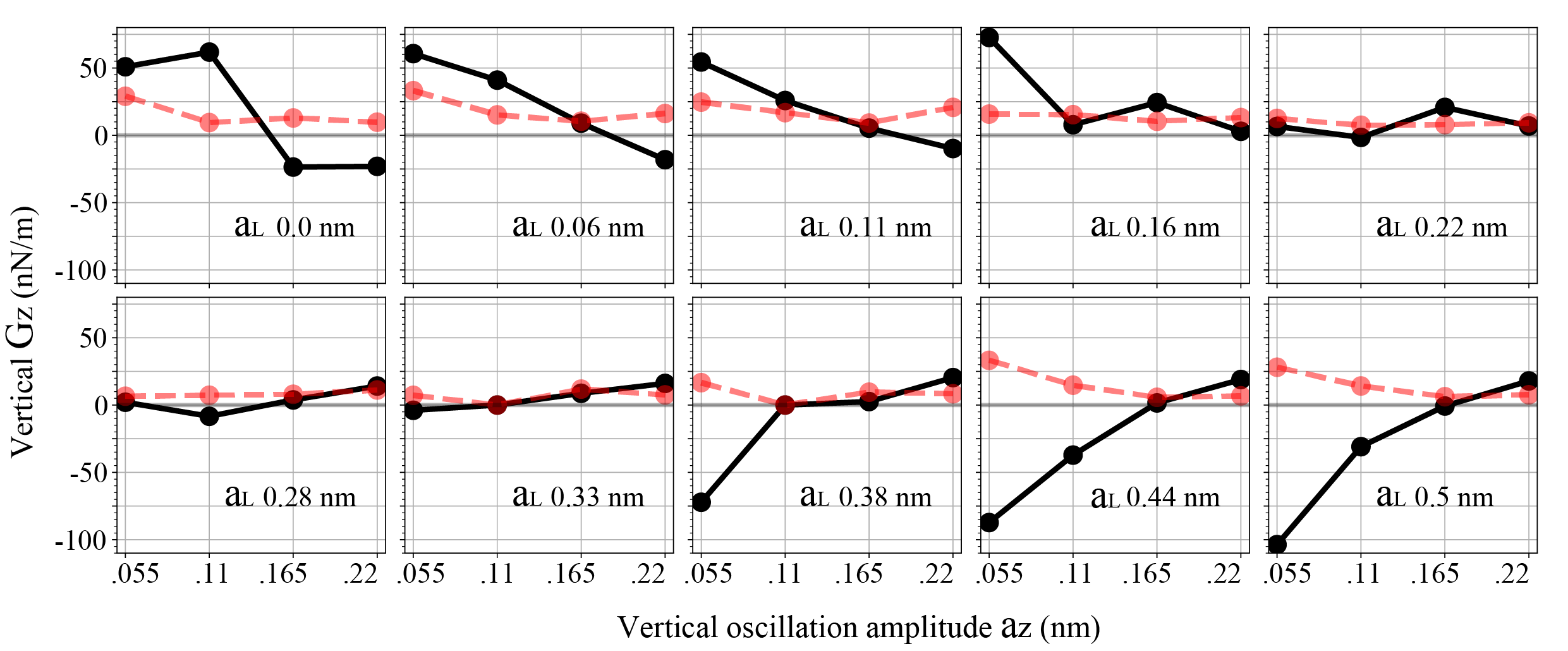}
\renewcommand\thefigure{S4}
\caption{Vertical $G$ as function of $a_z$, in the presence of lateral oscillation $a_L$. Black line is the real part of dynamical module, $G'$ and red is the imaginary part $G''$}\label{fig:S4}
\end{figure}

\end{document}